\documentclass[prb,aps,twocolumn,showpacs]{revtex4}

\usepackage{amssymb}
\usepackage{amsmath}
\usepackage{amsfonts}
\usepackage{tabularx}
\usepackage{pifont}
\usepackage{makeidx}
\usepackage[dvips]{graphicx}
\usepackage{graphics}
\usepackage{color}
\begin{document}
\bibliographystyle{apsrev}

\newcommand {\s}[3]{^{#1}\sigma_{#2}^{#3}}
\newcommand {\jpsj}{J. Phys. Soc. Japan\ }

\newcommand {\abs}[1]{| #1 |}
\newcommand {\bra}[1]{\langle \: #1 \: |}
\newcommand {\ket}[1]{| \: #1 \: \rangle}
\newcommand {\unity}{{\textrm{1}\hspace*{-0.55ex}\textrm{l}}}
\newcommand {\dotp}[3]{\langle \: #1 | \: #2 \: | \: #3 \: \rangle}
\newcommand {\expect}[1]{\langle \: #1 \: \rangle}
\newcommand {\ybco}[1]{YBa$_2$Cu$_3$O$_{#1}$}
\newcommand {\lacuo}{La$_2$CuO$_4$}
\newcommand {\cuo}{CuO$_2$}
\newcommand {\cuno}[2]{Cu$_{#1}$O$_{#2}$}
\newcommand {\Kk}[2]{^{#1}\!K^{#2}}
\newcommand {\K}[3]{^{#1}\!K_{#2}^{#3}}
\newcommand {\Kbar}[3]{^{#1}\overline{K}_{#2}^{#3}}
\newcommand {\R}[3]{^{#1}\!R_{#2}^{#3}}
\newcommand {\Rbar}[3]{^{#1}\overline{R}_{#2}^{#3}}
\newcommand {\ybc}{YBa$_2$Cu$_4$O$_8$}
\newcommand {\Cu}{$^{63}$Cu}
\newcommand {\Tone}{$^{63}$T$_1$}
\newcommand {\Tonec}{$^{63}$T$_{1c}^{-1}$}
\newcommand {\nd}{Nd$_2$CuO$_4$}
\newcommand {\ndce}{Nd$_{2-x}$Ce$_x$CuO$_4$}
\newcommand {\lacuosr}{La$_{2-x}$Sr$_x$CuO$_4$}
\newcommand {\infl}{SrCuO$_2$}
\newcommand {\infd}{Sr$_{1-x}$La$_x$CuO$_2$}
\newcommand {\srcl}{Sr$_2$CuO$_2$Cl$_2$}
\newcommand {\srf}{Sr$_2$CuO$_2$F$_2$}

\newcommand {\cuf}{Cu$_{5}$}
\newcommand {\cun}{Cu$_{9}$}
\newcommand {\cut}{Cu$_{13}$}

\title{Ab initio calculations of the electronic structure of cuprates using large scale cluster techniques}

\author{S. Renold, C. Bersier, E. P. Stoll,}
\email{estoll@gmx.ch}
\author{ P. F. Meier}

\affiliation{Physics Institute, University of Zurich, CH-8057 Zurich, 
         Switzerland}

\begin{abstract}
The local electronic structures of \lacuo, three members of the 
Yttrium-family (\ybco6, \ybco7, and \ybc ), and to some extent of \nd\ 
have been determined using 
all-electron ab-initio cluster calculations for clusters comprising up 
to thirteen planar copper atoms associated with their nearest planar 
and apical oxygen atoms. Spin-polarized calculations in the framework of 
density functional theory have enabled an estimation of the superexchange 
couplings $J$. Electric field gradients at the planar copper sites are
determined and their dependence on the occupation of the various atomic
orbitals are investigated in detail. The changes of the electronic field 
gradient and of the occupation of orbitals upon doping are studied and 
discussed. Furthermore, magnetic hyperfine fields are evaluated and 
disentangled into on-site and transferred contributions, and the chemical 
shifts at the copper nucleus are calculated. In general the results 
are in good agreement with values deduced from experiments except for the
value of the chemical shift with applied field perpendicular 
to the CuO$_2$-plane.
\end{abstract}
 

\maketitle

%************************************************************************
%************************************************************************

%{\bf 26.10.2007}

%**********************
%**********************
\section{Introduction}
\label{sec:intro}
%**********************
%**********************

The discovery~\cite{bib:bednorzmueller} of high temperature superconductivity
has initiated great experimental and theoretical efforts which aimed at a
detailed understanding of these materials. Nevertheless, a generally accepted
theory which explains at least the most important properties of the high
temperature superconductors could not yet be presented which is possibly 
related to the complex electronic structures of these materials.

Therefore, very early on ab-initio methods have been employed to determine the
electronic properties of these materials. Mostly, band-structure techniques
have been employed which are reviewed in Ref.~[\onlinecite{bib:Pickett}].
In our contrasting approach to electronic structure calculations, we use cluster
techniques in the framework of spin-polarized density functional theory with
localized basis functions which are especially well suited for the calculation
of local properties.

In particular, the cluster method has been successfully applied to the
determination of charge and spin-density distributions in the cuprate plane
and to a reasonably accurate evaluation of electric field gradients and
magnetic hyperfine fields for planar copper and oxygen nuclei in \lacuo\
and \ybco 7 using clusters comprising five planar copper and their nearest
neighboring planar and apical oxygen 
atoms~\cite{bib:huesser2000,bib:renold2001}.

In this work, these calculations are extended to include clusters
representative of \lacuo, \ybco 6, \ybco 7, \ybc, and to some extent of \nd. 
The development of both hard- and software has led to a significant
improvement of the quality of our calculations by use of larger
clusters comprising up to thirteen planar copper atoms. This allows a careful study of the convergence of the calculated local properties with respect to the cluster size.

The paper is organized as follows:
We will first outline the cluster technique and introduce the used clusters in
Sec.~\ref{sec:comput}. Sec.~\ref{sec:spin} is devoted to a discussion of the
spin distribution and contains estimations of exchange couplings in the
cuprates. In Sec.~\ref{sec:efg} we report on the calculation of electric field
gradients (EFG) at the planar copper sites in the different cuprates. 
The convergence of the results with respect to the cluster size is demonstrated and the various contributions of the molecular orbitals (MO) and atomic orbitals (AO) to the EFG are elucidated. The calculated distribution of charges and holes reveals that the 3$d_{3z^2-r^2}$ AO of the copper is not fully occupied. The doping dependence of these distributions and of the EFG are evaluated and compared to nuclear quadrupole resonance data. Sec.~\ref{sec:hyp} contains
a careful examination of hyperfine parameters and their disentanglement into 
on-site and transferred contributions. 
Calculations of chemical shieldings are presented in Sec.~\ref{sec:chem} and in particular, the role of the reference substance for magnetic shifts of the copper is discussed. The paper is terminated with the summary and conclusions in Sec.~\ref{sec:summary}. 

%*******************************************************
%*******************************************************
\section{The cluster technique}
\label{sec:comput}
%*******************************************************
%*******************************************************

The idea of the cluster technique is to select a contiguous region out of the
solid and to treat the electrons therein with standard many-body
theories. This so-called core region is surrounded by a shell of basis-free
pseudopotentials which prevent the electrons to be attracted by positive
point charges and provide smooth boundary conditions. The core and the boundary
regions are embedded in a large lattice of background point charges to ensure 
a very good approximation for the Madelung potential. In the case of the two 
undoped parent compounds, \lacuo\ and \ybco 6, the values chosen for the 
background point charges are based on the formal valence of the constituents.
In clusters representative of \ybco 7 and \ybc\ the formal valences
had to be slightly modified to reach charge neutrality.

The calculations were performed in the framework of spin-polarized density
functional theory which provides a good trade-off between accuracy and
computational cost. (Hartree-Fock theory, which completely neglects
correlation effects, yields poor results. In contrast, configuration
interaction methods, which correctly include both exchange and correlation 
effects in
their Hamiltonian, require enormous computer resources and are currently
prohibitive for the cluster sizes considered here.) For the representation of
the exchange and correlation functionals, the
potentials of Becke~\cite{bib:becke1,bib:becke2} and Lee, Yang, and
Parr~\cite{bib:LYP}, have been used. All atoms in the
core region of the clusters employ the standard triple zeta basis sets
(6-311G). For the determination of the ground state of the many-electron
system and the evaluation of the observed quantities, the Gaussian03 quantum
chemistry software package~\cite{bib:gaussian03} was used.

It is desirable to have as many atoms as possible in the core region, but the
available computer resources and the convergence of the self-consistent field
procedure are the limiting factors in this respect. As already mentioned, 
these limits have been pushed further since our first use of the cluster 
technique (see Ref.~[\onlinecite{bib:suter97}]) which now allows to use larger
clusters including up to nearly 1000 electrons.

In this work we present results for clusters comprising 5, 9, and
13 planar copper atoms together with their nearest planar and apical
oxygen atoms for clusters representative of \lacuo, \ybco 6, \ybco 7, \ybc, 
and \nd. In Table~\ref{tbl:clusters} the constitutive properties of all the 
used clusters for every substance are listed. The layout of the CuO$_2$ plane 
for the three cluster sizes is displayed in Fig.~2. The lattice constants
and the positions of the atoms in the unit cells for the three clusters of the
Y-family were chosen according to experimental structure determinations and 
are taken from
Refs.~[\onlinecite{bib:radaelli1994,bib:cava1990,bib:bordet1987,bib:fischer1989}].
The buckling of the planar oxygen atoms is not shown in Fig. 2.
For \lacuo, the calculations were performed for the tetragonal structure with 
lattice constants $a=b=3.77$~\AA, $c=13.18$~\AA\ and atomic positions according 
to Ref.~[\onlinecite{bib:radaelli1994}].

\begin{table}[htb]
\label{tbl:clusters}
\begin{tabular}{llrrrr} \hline
substance & name           & N & P & E & B \\ \hline
           & \cuno 5{26}    & 31 &  42 & 395 &  533 \\
\lacuo     & \cuno 9{42}    & 51 &  62 & 663 &  897 \\
           & \cuno {13}{62} & 75 &  78 & 991 & 1313 \\ \hline
           & \cuno 5{21}    & 26 & 194 & 345 &  468 \\
\ybco 6 & \cuno 9{33}    & 42 &  62 & 473 &  780 \\
           & \cuno {13}{49} & 62 &  86 & 841 & 1144 \\ \hline
           & \cuno 5{21}    & 26 &  37 & 345 &  468 \\
\ybco 7 & \cuno 9{33}    & 42 &  62 & 473 &  780 \\
           & \cuno {13}{49} & 62 &  86 & 841 & 1144 \\ \hline
           & \cuno 5{21}    & 26 &  37 & 345 &  468 \\
\ybc\ & \cuno 9{33}    & 42 &  53 & 473 &  780 \\
           & \cuno {13}{49} & 62 &  73 & 841 & 1144 \\ \hline
\end{tabular}
\caption{Compilation of the used clusters with their constitutive properties. N: number of atoms with a full basis set, P: number of atoms with (basis-free) pseudopotentials, E: number of electrons in the core region, B: number of basis functions.}
\end{table}

\begin{figure}[htb]
\includegraphics[width=0.5\columnwidth]{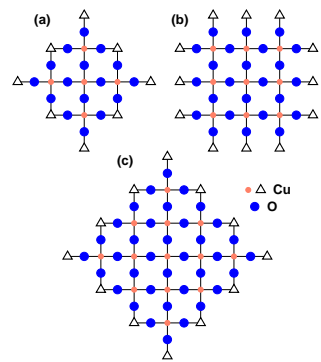}
\label{fig:cu5cu9cu13}
\caption{(color online). The layout of the CuO$_2$ planes in all of the used clusters, (a) with 5 planar copper atoms, (b) with 9 planar copper atoms, and (c) with 13 planar copper atoms. The apical oxygens are not shown. The empty triangles denote Cu$^{2+}$ ions simulated by a pseudopotential.}
\end{figure}

%**************************
%**************************
\section{Spin distribution}
\label{sec:spin}
%**************************
%**************************

Within the spin-polarized formalism, the spin multiplicity $M$ of a cluster is 
free parameter of the calculation. In a simple ionic picture, the planar
copper and oxygen atoms have a valence of
$+2$ and $-2$, respectively. This leads to a $3d^9$ configuration for the copper atom with a total
spin of one half whereas the oxygen valence gives zero
total spin. This suggests two choices of spin multiplicities with a physical
significance, a 
``ferromagnetic'' spin multiplicity with all spins parallel and an
``antiferromagnetic'' spin multiplicity with two neighboring spins being
antiparallel. However, other spin multiplicities are also possible leading to
spin alignments which can be viewed as superpositions of an antiferromagnetic
and a ferromagnetic spin state. We anticipate that the calculated total energy
is always lowest for $M$ that corresponds to an antiferromagnetic alignment. 
In Table~\ref{tbl:spinmult} the chosen spin
multiplicities $M$ are tabulated for each cluster size. The antiferromagnetic 
(ferromagnetic) spin multiplicities are in bold (normal) face. The 
superpositions are in parentheses.

\begin{table}[htb]
\begin{tabular}{ll}
cluster & multiplicities \\ \hline
\cuno 5{26}      & (2), {\bf 4}, 6 \\
\cuno 9{42}      & {\bf 2}, (4), (6), (8), 10 \\
\cuno {13}{62} & (4), {\bf 6}, (8), (10), (12), 14
\end{tabular}
\label{tbl:spinmult}
\caption{%\textcolor{red}
{Chosen spin multiplicities for the different clusters. Numbers in bold face denote multiplicities leading to an ``antiferromagnetic'' (4, 2, and 6) and a ``ferromagnetic'' (6, 10, and 14) spin alignment. Note that the possible spin multiplicities are determined solely by the number of planar copper atoms and are therefore independent on the specific material.}}
\end{table}

\begin{figure}[htb]
\includegraphics[width=0.5\textwidth]{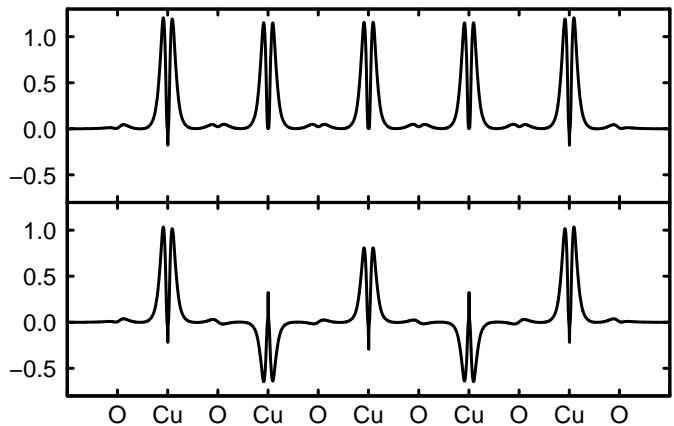}
\caption{Plot of the spin density along the planar Cu-O bonds in the Cu$_{13}$ cluster for \ybco 6. The upper (lower) panel shows the ferromagnetic (antiferromagnetic) spin configuration with $M=14$ ($M=6$).
}
\label{fig:spindensity}
\end{figure}

In Fig.~\ref{fig:spindensity} the spin density along the planar Cu-O bonds is
drawn for the large Cu$_{13}$ cluster representative of \ybco 6\ in the
case of $M=14$ corresponding to the ferromagnetic spin alignment (upper panel)
and in the case of $M=6$ corresponding to the antiferromagnetic spin alignment
(lower panel). The total energy for the latter spin multiplicity is 2.9 eV lower than that of the former. The double humps at the copper positions originate from the
approximately singly occupied $3d_{x^2-y^2}$ atomic orbital. The spins of 
neighboring coppers are indeed parallel for the ferromagnetic case and 
antiparallel for the antiferromagnetic case thus confirming the simple 
physical picture given above.
It is important to note that these spin density distributions are obtained also
in the smaller clusters and in the other substances considered.
A close inspection of the values at the copper sites (upper panel) shows a 
difference between the three inner coppers and the two coppers at the borders. 
This is due to the fact that the latter have only one nearest-neighbor (NN) Cu 
ion which implies a single transferred hyperfine field ($B>0$) whereas the 
formers have 4 NN leading to 4$B$. This will be further discussed in detail 
in Sec.~\protect \ref{sec:hypa}.

Furthermore, the ground state energy of the antiferromagnetic spin alignment is 
consistently lower than that of the ferromagnetic spin alignment. This important feature of our cluster method can be exploited to investigate on the exchange couplings in the different materials.

The Heisenberg Hamiltonian is given by
\begin{equation}
\mathcal H = -J \sum_{\expect{i,j}} \mathbf S_i \cdot \mathbf S_j
\end{equation}
where the summation is restricted to pairs of indices $(i,j)$ with the corresponding sites being nearest neighbored (and with $j>i$ to avoid double counting).

For a given material and cluster size, we can obtain ground state energies
$E_M$ and ground state wavefunctions $\psi_M(\mathbf r,\sigma) =
\phi_M(\mathbf r)\chi_M(\sigma)$ for any of the allowed multiplicities $M$ (see
Table~\ref{tbl:spinmult}). This is, however, not sufficient for a
determination of $J$ since $\dotp{\chi_M}{\sum \mathbf S_i \cdot \mathbf S_j}
{\chi_M}$ is in general not easy to determine. However, for a two spin state,
the Heisenberg model has an exact solution with $\dotp{\chi_3}{\mathbf S_1
  \cdot \mathbf S_2} {\chi_3} = 3/4$ and $\dotp{\chi_1}{\mathbf S_1 \cdot
  \mathbf S_2} {\chi_1} = -1/4$. This leads to the following idea:

We replace the expectation values of the spin operators in the Heisenberg
Hamiltonian by the product of the Mulliken spin densities at the respective
lattice sites,
\begin{equation}
\dotp{\chi_M}{\mathbf S_i \cdot \mathbf S_j}{\chi_M} = \rho_s^M(\textrm{Cu}_i)\rho_s^M(\textrm{Cu}_j),
\end{equation}
with the requirement that in the triplet state of the two-spin system
\begin{equation}
\alpha \rho_s^3 (\textrm{Cu}_i) \rho_s^3 (\textrm{Cu}_j) = 1
\end{equation}
\noindent
with a reduction factor $\alpha$.

(Note that the name ``spin density'' might be misleading. It is in fact not a spin density, but an integrated spin density, i.e. a spin. Nevertheless, we persevere with this term which is commonly used in quantum chemistry.)
With clusters comprising only two planar copper atoms, we determined $\rho_s^3(\textrm{Cu}_i) = 0.67$ for both copper sites which yields $\alpha = 2.23$. 
This reduction of the spin density from its value in an ionic model ($\rho_s^3(\textrm{Cu}_i) < 1$) is due to the fact that also the planar oxygen atoms carry a small amount of spin density. It has nothing to do with the reduction of the mean magnetic moment in the Heisenberg model by quantum fluctuations.
We thus arrive at the following modified Heisenberg-type equation for the determination of $J$:
\begin{equation}
E_M = -\alpha J \sum_{\expect{i,j}}\rho_s^M ( \textrm{Cu}_i ) \rho_s^M ( \textrm{Cu}_j ) \equiv -\alpha J \Sigma.
\label{eq:heisenberg}
\end{equation}

\begin{figure}[htb]
\includegraphics[width=0.95\columnwidth]{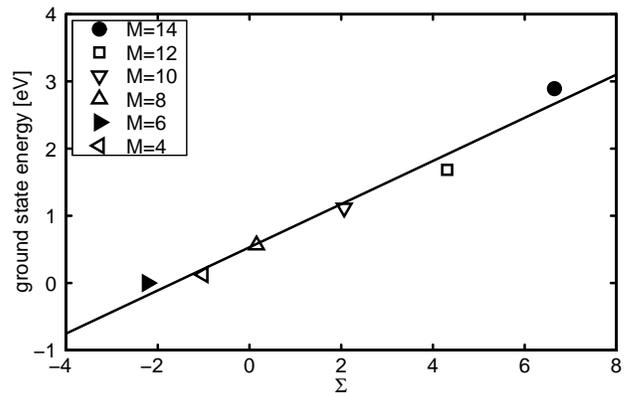}
\caption{Ground state energy versus $\Sigma$ (see Eq.~\ref{eq:heisenberg}) for 
all spin multiplicities $M$ of the Cu$_{13}$ cluster representative of \ybco 6.}
\label{fig:j_vs_e}
\end{figure}

In Fig.~\ref{fig:j_vs_e} the ground state energy is plotted versus $\Sigma$ for the different spin multiplicities $M$ in the case of the large Cu$_{13}$ cluster representative of \lacuo. A straight line with slope $\alpha J=350$~meV can be fitted. 
%However, it can also be argued that for the determination of the exchange couplings, only clusters which correspond to a simple physical picture should be considered, i.e. the slope in Fig.~\ref{fig:j_vs_e} should be determined considering only the two closed symbols which correspond to the antiferromagnetic and the ferromagnetic spin alignment.

The results of this fit are given in Fig.~%\ref{fig:exchangecouplings}.
4.
It is observed that the exchange couplings are fairly independent on the size of the cluster. Furthermore, they are all in the same range of about 150 meV. It is only the optimally doped \ybco 7 that has a slightly lower exchange coupling which is mainly due to the buckling of the planar oxygens. The two antiferromagnetic substances \lacuo\ and \ybco 6 and the underdoped \ybc\ have similar values of $J$.

The calculated antiferromagnetic couplings are in good agreement with data~\cite{keimer,shamoto} and also with theoretical values~\cite{munoz} obtained from smaller clusters but with more sophisticated ab-initio methods than used here.

We therefore conclude that the antiferromagnetic exchange interactions between two copper neighbors are an intrinsic property of the CuO$_2$ plane and depend only weakly on the specific material of experiments.

\begin{figure}[htb]
\includegraphics[width=0.95\columnwidth]{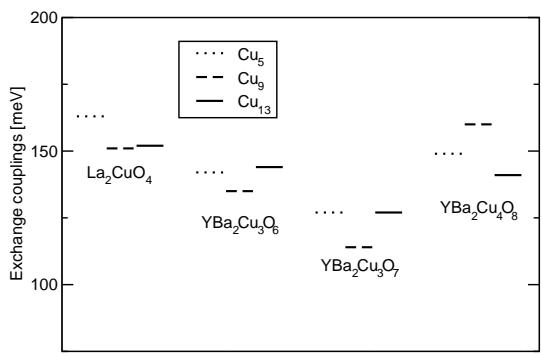}
\label{fig:exchangecouplings}
\caption{Exchange couplings fitted as described in the text.} % Dotted lines denote clusters containing 5 Cu atoms, dashed lines those with 9, and solid lines such with 13 Cu atoms.}
\end{figure}
\section{Electric field gradients}
\label{sec:efg}

\subsection{Outline}
\label{subsec:out}

Electric field gradients (EFG) are extremely sensitive to the (non-spherical)
charge distributions around the nucleus of interest. In the past twenty years a
large quantity of nuclear quadrupole resonance data has been accumulated for 
high-temperature superconducting cuprate compounds. Of particular interest 
are the EFGs at the planar copper site and their changes upon doping since they
reflect the local charge distribution and provide insight into the population 
of the different atomic orbitals.

We have already published calculated values for the EFGs in \lacuo,  \ybco7,
and \nd\ obtained for clusters with five copper atoms $(N=5)$ in the plane 
(see e.g. Refs.~[\onlinecite{bib:huesser2000,bib:renold2001}]). In this paper 
we add results for larger clusters ($N=9$ and 13) and report in addition results
for \ybco6 and \ybc. 

In subsection \ref{subsec:clusize} we first investigate on the influence of 
finite size effects. Next (subsection \ref{subsec:contr}) a detailed analysis of
the contributions to the copper EFG in terms of molecular orbitals is 
presented. We then postulate in subsection \ref{subsec:approx} an approximate 
relationship between the copper EFG values and the partial atomic Mulliken 
charge populations of the orbitals $3d_{x^2-y^2}$ and $3d_{3z^2-r^2}$.
Although approximate within about 10 \% only, this provides insight into the
EFG differences observed in different compounds as well as their changes 
upon electron or hole doping. The latter relationship is discussed in 
subsection \ref{subsec:dopdep}.

In subsection \ref{subsec:compar} the theoretical results are compared to 
experimental data and an extended discussion about the interpretation of the 
results is presented.

\subsection{Dependence of the calculated Cu EFG on the cluster size}
\label{subsec:clusize}

%In Fig. \ref{fig:Vzz} we present the main component $V_{zz}$ of the EFG 
In Fig. 5 we present the main component $V_{zz}$ of the EFG
obtained for the copper in the center of clusters comprising $N$ = 5, 9, and 13 
planar copper atoms simulating the \lacuo,\,  \ybc,\, \ybco6, and \nd\ compounds. 
These values have been calculated for spin multiplicity $M$ = 4, 2, and 6, 
respectively, which correspond to an antiferromagnetic alignment of the spins 
and yield the lowest total energy.  The values for $N$ = 9 and 13 are nearly 
equal while those for $N$ = 5 are about 10  \% to 20 \% smaller due to finite size 
effects. We note that the EFG values for the \ybco7 compounds exhibit the same 
size dependence with values for \ybco7 between the those for \lacuo\  and 
\ybco6 (see subsection \ref{subsec:dopdep}). 

\begin{figure}[htb]
\includegraphics[width=1.05\columnwidth]{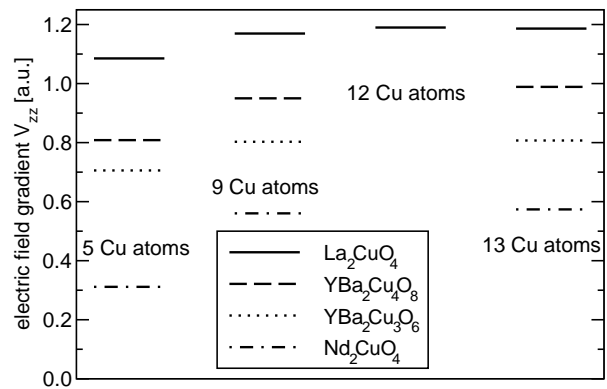}
\label{fig:Vzz}
\caption{EFG value $V_{zz}$ as a function of the number of Cu atoms in 
the clusters.} %for \lacuo\  (solid line), \ybc\  (dashed line), \ybco6
%(dotted line), and \nd\ (dashed-dotted line).}
\end{figure}

In addition, we have calculated the EFG value for a La$_2$CuO$_4$ cluster with
an even number ($N$ = 12) of copper atoms without spin polarization (singlet 
state with $M$ = 1). As seen in Fig. %\ref{fig:Vzz} this value is very close 
5 this value is very close to those calculated for $M$ = 9 and 13.

We further note that $V_{zz}$ is given in atomic units (1~a.u. corresponds to
9.71525 $\times$ 10$^{21}$ V m$^{-2}$). A comparison to experiments will be 
made in subsection \ref{subsec:compar}.

%%%%%%%%%%%%%%%%%%%%%%%%%%%%%%%%%%%%%%%%%%%%%%%%%%%%%%%%%%%%%%%%%%%%%
\subsection{Detailed analysis of contributions}
\label{subsec:contr}
%%%%%%%%%%%%%%%%%%%%%%%%%%%%%%%%%%%%%%%%%%%%%%%%%%%%%%%%%%%%%%%%%%%%%

In this subsection we first analyze the various contributions to the 
copper EFG in detail. It will be shown that the values of the EFG are not 
solely determined by the occupancies of the $3d_{x^2-y^2}$ atomic orbital 
as is often assumed in simplistic models.

The charge distribution is determined by the occupied MOs. The $m^{th}$ 
MO is represented as a linear combination \cite{footnote}

\begin{equation}
\phi_m(\vec{r})=\sum_{K=1}^n\phi_m^K(\vec{r}-\vec{R}_K)=\sum_{K=1}^n\sum_{k=1}^{n_K}c_m^{K,k}B_{K,k}(\vec{r}-\vec{R}_K)
\label{eq:linearcomb}
\end{equation}
of $n_K$ atomic basis functions $B_{K,k}$ centered at the nuclear sites 
$K=1,\dots,n$, and the $c_m^{K,k}$ are the MO coefficients.
We assume in the following that the target nucleus $K_T$ is at 
$\vec{R}_{K_T}=0$. The contribution of the MO $\phi_m$ to the EFG at 
$K_T$ is given by the matrix element
\begin{equation}
V^{(m)}_{ij}=\dotp{\phi_m(\vec{r})}{\frac{3x^ix^j-r^2\delta^{ij}}{r^5}}{\phi_m(\vec{r})}
\nonumber
\end{equation}
\begin{equation}
=\sum_{K=1}^{n}\sum_{L=1}^{n}\sum_{k=1}^{n_K}\sum_{l=1}^{n_L}c_m^{K,k}c_m^{L,l} 
\times
\label{eq:matrixel}
\nonumber
\end{equation}
\begin{equation}
\times 
\dotp{B_{K,k}(\vec{r}-\vec{R}_K)}{\frac{3x^ix^j-r^2\delta^{ij}}{r^5}}{B_{L,l}(\vec{r}-\vec{R}_L)}.
\end{equation}

This matrix element contains contributions from basis functions centered 
at two nuclear sites $K$ and $L$. Thus we can identify three types of 
contributions: $(i)$ on-site terms from basis functions centered at the 
target nucleus ($K=L=K_T$, contribution I), $(ii)$ mixed on-site 
off-site contributions (II), and $(iii)$ purely off-site terms with 
$K\neq K_T$ and $L \neq K_T$ (III).

In addition, there is a contribution coming from all nuclear charges $Z_K$ and all point charges mentioned in Sec. II which we denote by 
\begin{equation}
W_{ij}^{\rm tot} = W_{ij} +  W_{ij}^{\rm pc}
\end{equation}
where W$_{ij}$ refers to nuclei in the core region of the
cluster and  $W_{ij}^{\rm pc}$ to the point charges outside this core.
The charges of the bare 
nuclei at sites $K \ne K_T$ in the core are screened by the matrix 
elements of contribution III with $K 
= L \ne K_T$. Therefore the combined contributions from III and the nuclei (W$_{ij}$) are small. The partitioning of the contributions to the EFG tensor $V_{ij}$ thus reads
\begin{equation}
V_{ij}={^IV_{ij}}+{^{II}V_{ij}}+{^{III}V_{ij}}+W_{ij} + W_{ij}^{\rm pc}.
\end{equation}
 More details about these regional partitions can be found in 
Ref.~[\onlinecite{bib:stoll}].

In Table~\ref{tab:contr} the contributions to the EFG 
component V$_{zz}$ for the copper in the center of the Cu$_{13}$ 
clusters representative for Nd$_2$CuO$_4$, YBa$_2$Cu$_3$O$_7$, and 
La$_2$CuO$_4$ are collected. As expected, the values from region III and 
W as well as those from point charges outside the core region are 
small. However, mixed on-site off-site contributions (region II) are 
substantial. They are mainly transferred via the on-site 3$d_{3z^2-r^2}$
orbital which has a non-negligible overlap with the 2$p_z$ orbitals of 
the four surrounding planar oxygens.
The on-site terms (region I) contain values which come 
from terms in the evaluation of matrix elements (\ref{eq:matrixel}) where one 
basis function is $s$ like and the other $d$ like. Their contribution is denoted
by R in Table V. Also the contributions 
from the Cu $p$ type orbitals are substantial due to the large 
$\expect{r^{-3}}$ values.
The occupancies of the three $d$ orbitals with t$_{2g}$ symmetry are close 
to 2 so that their combined contribution to $V_{zz}$ is small. As 
expected, the distinguished AO is the $3d_{x^2-y^2}$ with a polarization of
70 \% accompanied with smaller 
but still significant polarization in the orbital $3d_{3z^2-r^2}$.

\subsection{Model}
\label{subsec:approx}

As pointed out above, a rigorous explanation for the variation of the copper 
EFG values in different cuprates is very complicated. A simple explanation, 
however, which concentrates on the two most relevant orbitals, is
given in this subsection.

\begin{figure}[htb]
\includegraphics[width=0.48\textwidth]{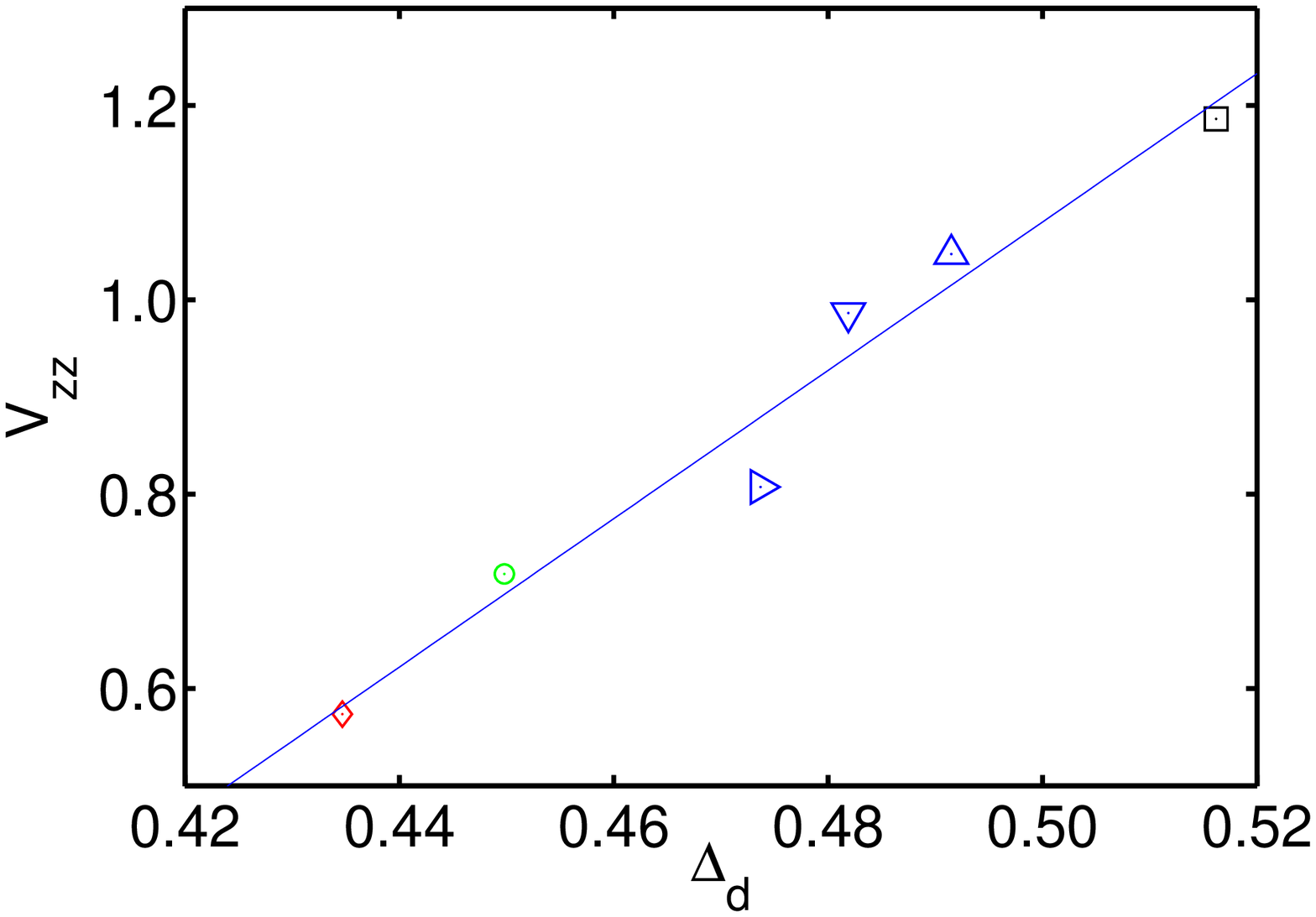}
\caption{(color online). Theoretical EFG component $V_{zz}$ versus $\Delta_d$=$p_c$(3d$_{3z^2-r^2}$)$-p_c$(3d$_{x^2-y^2}$) for \lacuo\ (black square), \ybco{7} (blue triangle up), \ybc\ (blue triangle down), \ybco{6} (blue triangle right), \srcl\ 
(from Ref~[\protect \onlinecite{bersier2002}]) (green circle), and \nd\ (red diamond). The straight line is a fit $V_{zz}=7.63\cdot\Delta_d-2.74$.}
\label{fig:EFG_deltad}
\end{figure}

Previously, we have investigated in Refs.~[\onlinecite{bib:stoll1,bib:bersier}] 
the changes of the copper EFG 
values that occur upon doping in La$_2$CuO$_4$ and in Nd$_2$CuO$_4$. We 
found that it is sufficient to concentrate on those MOs which contain 
partially occupied $3d_{x^2-y^2}$ and $3d_{3z^2-r^2}$ AOs at the target 
nucleus. The EFG is then given by
\begin{equation}
V_{zz}=S+\frac{4}{7} \left[N_{3z^2-r^2}\langle r^{-3} 
\rangle_{3z^2-r^2}-N_{x^2-y^2} \langle r^{-3} \rangle_{x^2-y^2} \right]
\label{eq:Vdelta}
\end{equation}
where $S$ is the contribution from all other orbitals and regional 
partitions. The values for $\langle r^{-3} \rangle$ calculated for the 
different 3$d$ atomic orbitals are almost identical (e.g. $\langle r^{-3} 
\rangle_{3z^2-r^2} = 8.004$ a.u. and $\langle r^{-3} \rangle_{x^2-y^2} = 8.042$
a.u.) and therefore are replaced by the average $<r^{-3}> = 8.020$ a.u.. We 
further note that the partial occupation numbers $N$ of the AOs 
are very similar to the partial Mulliken populations $p_c$ which are 
gathered in Table~\ref{tab:contr2}. In particular, the differences
\begin{equation}
\Delta_d=p_c(3d_{3z^2-r^2})-p_c(3d_{x^2-y^2})
\label{eq:deltad}
\end{equation}
and
\begin{equation}
N_{3z^2-r^2}-N_{x^2-y^2}
\end{equation}
are very similar.
In Fig.~\ref{fig:EFG_deltad} we therefore plot the calculated copper EFG 
component $V_{zz}$ versus $\Delta_d$. The straight line corresponds to
\begin{equation}
V_{zz}=-2.74+7.63 \Delta_d 
\label{eq:deltadV}
\end{equation}
\noindent
which emphasizes the quality of the model.
\begin{table}
\caption{ Contributions to the EFG component $V_{zz}$ for the copper in the center of the Cu$_{13}$ clusters representative for Nd$_2$CuO$_4$, YBa$_2$Cu$_3$O$_7$ and La$_2$CuO$_4$.
  }

\begin{tabular}{lrrr}%|rr}
\hline
                           & Nd$_2$CuO$_4$&YBa$_2$Cu$_3$O$_7$&La$_2$CuO$_4$\\ \hline%& La h.d. M5& Nd e.d. M5   \\ \hline
W$^{pc}$                   & $-0.033$  &  0.021 & 0.019  \\ \hline%& 0.021        & $-0.033$     \\ \hline
I &&& \\ \hline%&& \\
R                  &   0.755  & 0.640 & 0.474 \\ \hline%& 0.619        &   0.797        \\
p                          & $-1.431$  &$-1.157$ & $-1.250$ \\ \hline%& $-1.132$       & $-1.433$      \\
d$_{x^{2}-y^{2}}$          & $-6.378$  &$-6.344$ & $-6.263$ \\  
% \hline & $-6.294$       & $-6.613$       \\
d$_{xy}$+d$_{xz}$+d$_{yz}$ &  $-0.072$ &$-0.141$ & $-0.090$ \\ 
% \hline & $-0.153$       & $-0.064$       \\
d$_{3z^{2}-r^{2}}$         &   8.309   & 8.630 &  8.736 \\ \hline%& 8.660        &   8.222        \\ \hline
II &&& \\ \hline%&& \\
s                          & $-0.003$  &$-0.004$ & 0.004 \\ \hline%&$-0.004$        & $-0.002$       \\
p                          & $ 0.013$   & 0.014&  0.022 \\ \hline%& 0.014        & $ 0.013$       \\
d$_{x^{2}-y^{2}}$          & $-0.021$  &$-0.020$ & $-0.019$ \\ 
%\hline &-0.019        & $-0.020$       \\
d$_{xy}$+d$_{xz}$+d$_{yz}$ &  $ 0.001$  & 0.001 & 0.002 \\ 
%\hline & 0.001        & $ 0.001$       \\
d$_{3z^{2}-r^{2}}$         & $-0.585$  &$-0.515$ & $-3.382$ \\ \hline%&$-0.498$        & $-0.616$       \\ \hline
III + Nuclei               & $-0.015$  &$-0.061$ & $-0.053$ \\ \hline%&$-0.060$        & $-0.017$        \\ \hline
Total                      &  0.538     & 1.066 & 1.202 \\ \hline%&  0.234       &   0.768         \\
\end{tabular}
\label{tab:contr}

\end{table}

\begin{table}
\caption{Partial occupation numbers $N$ and partial Mulliken populations $p_c$
for the frontier orbitals $3d_{x^2-y^2}$ and $3d_{3z^2-r^2}$.
  }
\begin{tabular}{lrrr}
                           & Nd$_2$CuO$_4$&YBa$_2$Cu$_3$O$_7$&La$_2$CuO$_4$ \\ \hline
$N(x^2-y^2)$               & 1.379         & 1.375  & 1.363 \\
$p_c(x^2-y^2)$             & 1.430         & 1.423  & 1.406 \\
$N(3z^2-r^2)$              & 1.820         & 1.898  & 1.910 \\
$p_c(3z^2-r^2)$             & 1.865        & 1.914  & 1.922 \\

\end{tabular}
\label{tab:contr2}
\end{table}
%%%%%%%%%%%%%%%%%%%%%%%%%%%%%%%%%%%%%%%%%%%%%%%%%%%%%%%%%%%%%%%%%%%%%%%%%%%%%%%%
\subsection{Charge and hole distribution and qualitative discussion of bonding}
\label{subsec:charghol}
%%%%%%%%%%%%%%%%%%%%%%%%%%%%%%%%%%%%%%%%%%%%%%%%%%%%%%%%%%%%%%%%%%%%%%%%%%%%%%%%

The atomic Mulliken charges $\rho$ for the central copper and neighboring 
oxygen atom in \lacuo\ as calculated for the Cu$_{13}$ with spin multiplicity
$M = 6$ are
\begin{equation}
\rho_0({\rm Cu})= 1.143, \rho_0({\rm O}_p)=-1.635, \rho_0({\rm O}_a)=-1.945 .
\end{equation}
\noindent
The corresponding values for the four units surrounding the central one are
\begin{equation}
\rho_1({\rm Cu})= 1.131, \rho_1({\rm O}_p)=-1.629, \rho_1({\rm O}_a)=-1.944 .
\end{equation}
\noindent
while those of units at the boundary of the core region differ at most by
one percent due to finite size effects.

We define
\begin{equation}
\rho_0 (3) = \rho (Cu) + 2 \rho (O_p) + 2 \rho (O_a) + 6 
\end{equation}
\noindent
where 6 accounts for the charges of the two La ions in the unit cell. We
obtain $\rho_0 (3) = -0.012$ that is close to zero which emphasizes
the suitability of these clusters to represent the local
conditions of atoms in the crystal. This point cannot be overstressed since
it also means that there is now also confidence in the partial Mulliken
populations of the individual orbitals which have been studied in detail
elsewhere [\onlinecite{bib:stoll2}]. 

The partial Mulliken populations of those AO which differ from 2 by more than
0.02 are given in Table~\ref{tab:partmul} and are visualized in 
%Fig.~\ref{fig:popu} where the area in yellow (gray on top) accounts for an 
Fig.~7 where the area in yellow (gray on top) accounts for an intrinsic hole 
of 1.5 missing electrons which is compensated by the occupancy of the 4$s$ 
orbital. All attempts to deduce the distribution of holes which do not take into
account the $4s$ orbital are very questionable and misleading.

\begin{figure}[htb]
\includegraphics[width=0.5\columnwidth]{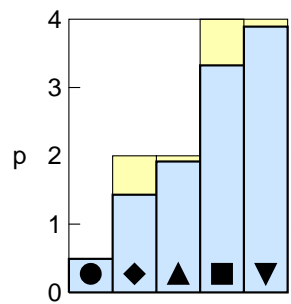}
\label{fig:popu}
\caption{(color online). Partial Mulliken populations of the atomic orbitals 
(tinted in darker gray (blue)) for La$_2$CuO$_4$: Black symbols denote 
$4s$ (circles), $3d_{x^2-y^2}$ (diamonds), and $3d_{3z^2-r^2}$ (triangles up), 
2 $\times$ $p($O$_p(2p_{\sigma}))$ for the planar oxygens (squares), and 2
$\times$ $p($O$_a(2p_{z}))$ for the apical oxygens (triangles down).
The light gray (yellow) tinted areas denote the missing charge against the 
simple ionic model.}
\end{figure}

\begin{table}
\caption{Partial Mulliken populations $p_c$ of the AOs in the central unit
of \lacuo. }
\begin{tabular}{l|rrrrr}
AOs\, &\,3$d_{x^2-y^2}$& 3$d_{3z^2-r^2}$& 4$s$& 2$p_\sigma(O_p)$ & 2$p_z(O_a)$ \\
\hline 
$p_c$&\,\,\,\,\, 1.406&\,\,\,\,\, 1.922&\,\,\,\,\, 0.491&\,\,\,\,\, 1.662&\,\,\,\,\, 1.946 \\ \hline
\end{tabular}
\label{tab:partmul}
\end{table}
The ionic picture which assigns charges of +3(La), +2(Sr), +2(Cu), and $-2$(O), respectively is a reasonable approximation
for the out-of-plane La- and apical O$_a$-atoms. The copper and the planar
oxygen atoms, however, are rather covalently bound and the relation between
charge and hole transfer is not trivial. A Mulliken charge population
analysis~[\onlinecite{bib:stoll2}] attributes a charge of 1.16 to the copper and
$-1.64$ to the planar oxygens but 40\% of the hole is transferred to the oxygens
leaving 60\% on the copper. Since a hole transfer is accompanied by an electron
charge transfer in the opposite direction it is concluded that of the total
of 0.84 electrons transferred from the oxygen to the copper, 0.40 electrons are
accounted for the hole transfer. The remaining 0.44 electrons almost
correspond exactly to the Mulliken 4s orbital population. The discrepancy of
0.05 electrons can be attributed to secondary interactions which also involve
the 3$d_{3z^2-r^2}$ and the 2$p_z$ apical oxygen orbitals. The substantial
occupation of the Cu 4$s$ orbital is responsible for the existence of the
hyperfine field transferred from the four next nearest copper atoms which
is revealed by NMR experiments.
%%%%%%%%%%%%%%%%%%%%%%%%%%%%%%%%%%%%%%%%%%%%%%%%%%%%%%%%%%%%%%%%%%%%%%%%%%%%%%%%
\subsection{Doping dependence}
\label{subsec:dopdep}
%%%%%%%%%%%%%%%%%%%%%%%%%%%%%%%%%%%%%%%%%%%%%%%%%%%%%%%%%%%%%%%%%%%%%%%%%%%%%%%

The measured copper quadrupole frequencies generally increase with hole doping. In particular, in La$_{2-x}$Sr$_x$CuO$_4$, the slope of this increase is reported in~\cite{bib:japaner,bib:imai1993,bib:haase} to be in the range of 0.56$x$ to 0.7$x$. It is of utmost interest to deduce the redistribution of charges that occurs upon doping from these data. This is, however, a complex task because the doped holes will also change the lattice parameters and atomic positions in the unit cell which in turn influence the EFG.

Ab-initio calculations of doping-induced charge redistribution have been reported by Ambrosch-Draxl {\it et al.}~[\onlinecite{Ambrosch}] for the compound HgBa$_{2}$CuO$_{4+\delta}$. They employed the full-potential linearized augmented plane-wave method and used a series of supercells containing one excess oxygen atom. In principle, total-energy and atomic-force calculations can also be performed for small clusters. They are not feasible, however, for large clusters. We therefore report in the following on investigations of doping dependence with lattice parameters and atomic positions kept fixed at values corresponding to the undoped compound La$_2$CuO$_4$. The results are therefore only reliable for small doping, i.e. they should be considered to describe the (linear) slope of changes around the undoped material. We will turn back to changes in the atomic positions at the end of this subsection.

The method of cluster calculations would also enable the study of changes in the local electronic structure that occur by replacing e.g. a trivalent La by a bivalent Sr. These inhomogeneous changes, however, are not the subject of the present investigation where we simulate the long range effects of doping (delocalized holes or electrons) by two different approaches. First we introduce additional point charges at the periphery of the cluster to place an electric field across the cluster to move charges toward or away from the atoms of interest in the cluster center. In this so called ``peripheral charges method'', the added system of charges has no physical interpretation except that it can be continuously altered so that charge can be progressively directed to or extracted from important ions of interest in the cluster. The cluster and the system of charges in total keep the same number of electrons and spins but, using for instance a Mulliken population analysis approach, the charge of the cluster center can be progressively changed in a manner expected by doping.
In the second approach of simulating doping we have added or removed two electrons from the cluster and repeated the calculation. Changing the number of electrons by an even number allows us to keep the spin state.

\begin{figure}[htb]
\includegraphics[width=0.48\textwidth]{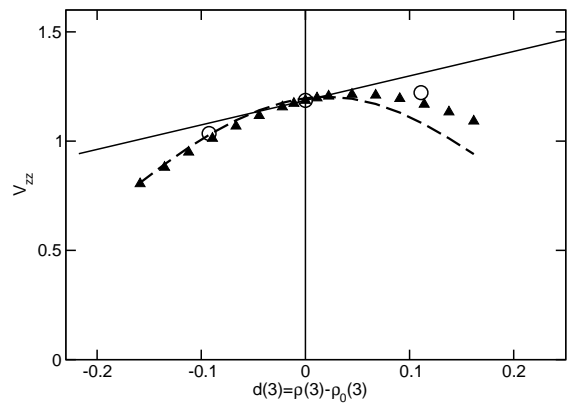}
\caption{$V_{zz}$ versus ''doping level'' $d(3)=\rho(3)-\rho_0(3)$ for
\lacuo.}
\label{fig:doping}
\end{figure}

\begin{figure}[htb]
\includegraphics[width=0.48\textwidth]{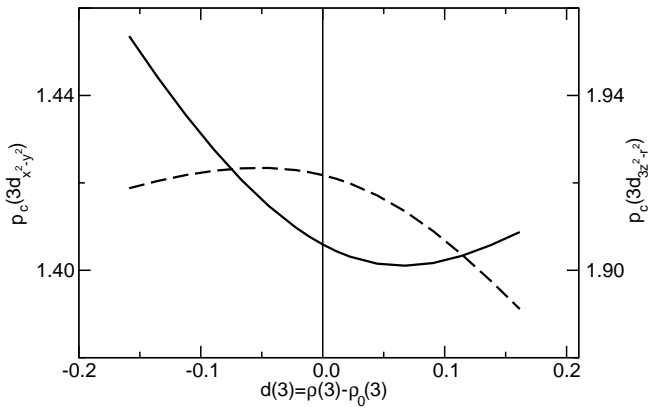}
\caption{$p_c(3d_{x^2-y^2})$ (solid curve) and $p_c(3d_{3z^2-r^2})$ (dashed
curve) versus ''doping level'' $d(3)=\rho(3)-\rho_0(3)$ for \lacuo. }
\label{fig:pdzx_La}
\end{figure}

To discuss the EFGs for doped \lacuo\ we define 
\begin{equation}
\rho (3) = \rho ({\rm Cu}) + 2 \rho ({\rm O}_p) + 2 \rho ({\rm O}_a) + 2 \rho({\rm La}),
\end{equation}
\noindent
where the Mulliken charges refer to the central unit and $\rho({\rm La}) = 3$.

For the undoped Cu$_{13}$ cluster with spin multiplicity $M$=6 we obtain 
$\rho_0(3) = -0.012$ as discussed above. In Fig.~\ref{fig:doping} we plotted 
the EFG component $V_{zz}$ for the central copper versus the doping level
$d(3) = \rho (3) - \rho_0(3)$. Note that negative values of $d$ would apply to 
electron-doped materials which for the present case of \lacuo\ is of course 
of no experimental relevance. The two circles at $d=-0.09$ and $d=0.11$
have been obtained by adding and subtracting two electrons, respectively. The 
triangles denote results obtained with the peripheral charge method 
[\onlinecite{bib:stoll1}]. All values have been calculated for 
multiplicity $M$=6.

The slope of the increase of $V_{zz}$ at $d = 0 $ is 1.12. An analysis of the populations of the frontier orbitals is shown in Fig.~\ref{fig:pdzx_La}. For small $d(3)$ but with increasing (hole) 
doping $p_c(3d_{3z^2-r^2})$ is increased but reaches a maximum at $d(3)=-0.056$ 
while $p_c(3d_{x^2-y^2})$ is decreased and has a minimum at 0.066. Using 
Eqs. (\ref{eq:Vdelta}), (\ref{eq:deltad}) and (\ref{eq:deltadV}), the dashed
line in Fig.~\ref{fig:doping} shows the calculated $V_{zz}$. 
 
An analysis of the charges in the occupancies upon doping exhibits that an additional extrinsic hole goes to 15 \% to the 3$d_{x^2-y^2}$ AO, to $2 \times 18.7 \, \% = 37.4 \, \%$ to the two O$_p$ $2p_{\sigma}$ AO, to 7 \%  to the 3$d_{3z^2-r^2}$ AO and to $2 \times 18.7 \, \% = 37.4 \, \%$ to the two O$_a$ $2p_{z}$ AO, while the $4s$ remains practically constant.

It is instructive to compare these results with those obtained in Ref. [\onlinecite{Ambrosch}] for HgBa$_{2}$CuO$_{4+\delta}$. Although the calculational procedures are quite different the essential conclusions are the same. We first define

\begin{equation}
\rho(2)=\rho(3d_{x^2-y^2})+2\rho(2p_{\sigma})
\end{equation}
as the charge of the planar orbitals and correspondingly $d(2)=\rho(2)-\rho_0(2)$ as the derivation from the undoped case. We get

\begin{equation}
\rho(2)=0.52 \, \rho(3)
\end{equation}

\noindent
which is to be compared with $\rho(2)=0.55 \, \rho(3)$ in Ref. [\onlinecite{Ambrosch}]. Thus in both cases the removal of an electron by a dopant atom in the intra-layer induces only half a hole in the CuO$_2$ plane. The calculated changes of the occupancies of the individual orbitals are somewhat different. For HgBa$_{2}$CuO$_{4+\delta}$ a decrease of 35 \% for $p_c(3d_{x^2-y^2})$ was reported whereas our value for La$_2$Cu$_3$O$_4$ is 15 \%. This is compensated by a smaller decrease of 20 \% for $2 \times p_c(2p_{\sigma})$ compared to 37 \%. These differences are, however, of not too much relevance since the assignment of the charge to AO in covalent bonds is anyhow somewhat arbitrary.
The turn-over of $V_{zz}$ in Fig. 8 and in the occupancies in Fig. 9 may well be an artefact of the not appropriately adjusted changes in the atomic positions.
It should be pointed out, however, that in the calculations of Ambrosch-Draxl {\it et al.} [\onlinecite{Ambrosch}] where these positions have been adjusted, the initially linear changes stop at a doping concentration $d(3)=0.22$ and remain constant at higher $d(3)$ values which indicates a saturation of the planar hole content at 0.12.

The essential question now is whether these theoretically predicted redistributions of charges can be corroborated by experimental facts. The slope of increase of $V_{zz}$ of 1.1 is too large compared to the data. We may explain this partially with arguing that the lattice parameter $a$ shrinks upon doping. While it is not feasible to determine the ground-state energy of all atoms with cluster calculations, it is straightforward to study the changes if a single parameter is varied and the results are reliable since relative changes are involved. We have previously reported~\cite{bib:renold2003a} on the changes that occur when the lattice parameter $a$ for a cluster representative of La$_2$CuO$_4$ with 5 atoms is varied and found that when $a$ shrinks by 1~\% the EFG value $V_{zz}$ is reduced by $\approx 10.7$~\%. Assuming a reduction of $a$ by 4 \% upon doping, according to data presented in Ref. [\onlinecite{bib:bozin}], we get a reduced slope of $V_{zz}$ of $1.12-0.43=0.69$, which is (very) close to the experimental value.  Notice, however, that this argumentation neglects the variation of $V_{zz}$ due to small changes of the positions of the atoms in the $c$-direction, in particular those of the apex oxygen and of the La ions.

\subsection{Comparison with NQR experiments}
\label{subsec:compar}

For a nuclear spin 3/2, the connection between NQR frequencies $^{63}\nu_Q$ and
the main component $V_{zz}$ of field gradients  in the case of axial symmetry
is given by\cite{slichter}:

\begin{equation}
^{63}\nu_Q = \frac{\Delta E(\pm 3/2 \rightarrow \pm 1/2)}{h} = \frac{e\ ^{63}Q V_{zz}}{2h}
\end{equation}
with $^{63}Q$ being the nuclear quadrupole moment. Unfortunately, directly 
measured values for $^{63}Q$ are not available. The commonly used value of 
$^{63}Q = -0.211 \pm 0.004$~b was obtained some time ago by 
Sternheimer\cite{bib:pyykkoe2001} by interpreting excitation spectra with the 
Hartree-Fock approximation.

\begin{figure}[htb]
\includegraphics[width=0.48\textwidth]{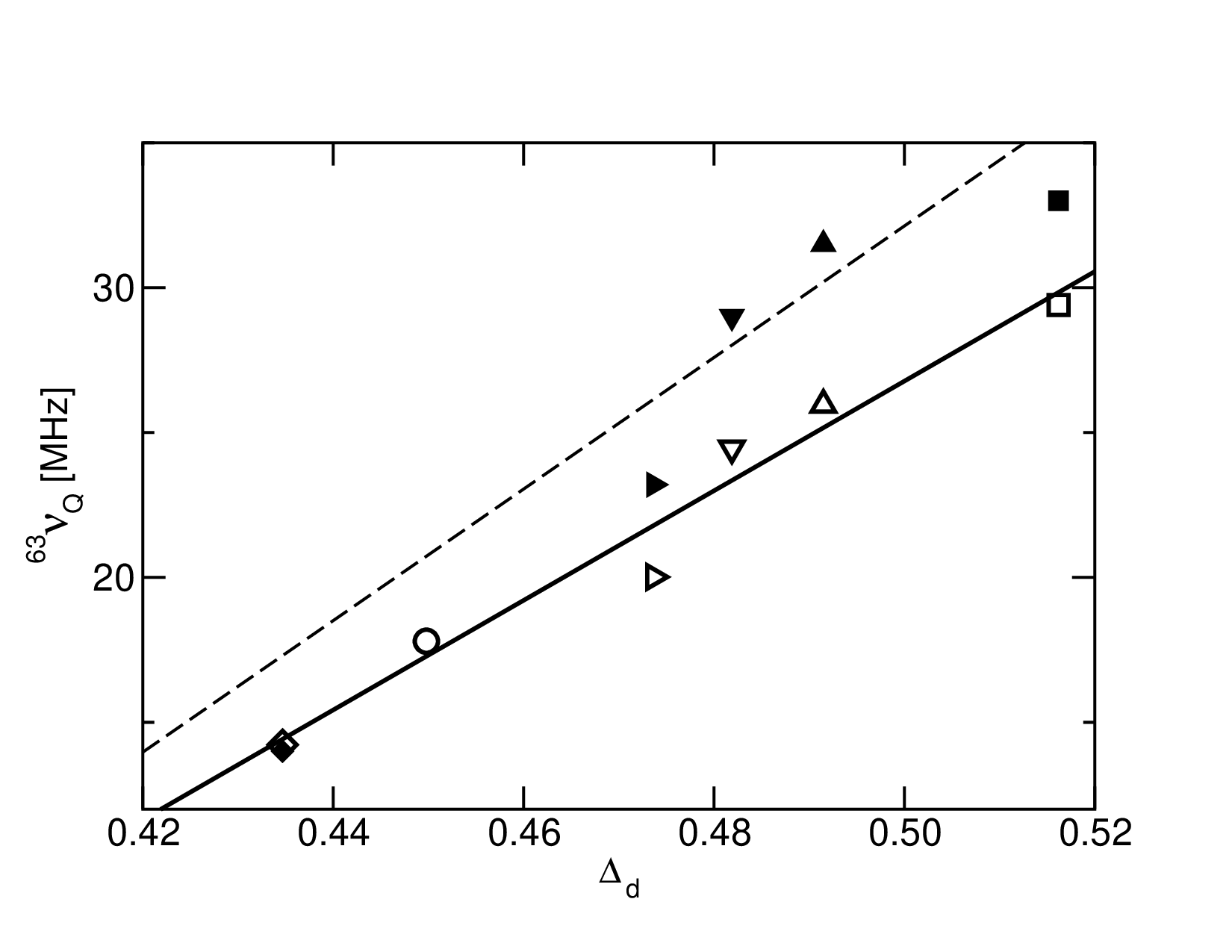}
\caption{Experimental %~\protect \cite{matsumura}
(full symbols) and calculated (open symbols) quadrupole
frequencies $^{63}\nu_Q$ as a function of the calculated difference of the
Mulliken populations $\Delta_d$. The symbols denote: \lacuo\ (square), \ybco{7} (triangle up), \ybc\ (triangle down), \ybco{6} (triangle right), \srcl\
(from Ref~[\protect \onlinecite{bersier2002}]) (circle), and \nd\ (diamond).
}
\label{fig:theoex}
\end{figure}
In Fig.~\ref{fig:theoex} we plot our theoretical values for the quadrupolar 
frequency $^{63}\nu_Q$ obtained from the calculated values $V_{zz}$ with the 
above mentioned value for $^{63}Q$ versus the calculated difference of the
Mulliken populations $\Delta_d$. The straight line corresponds to

\begin{equation}
^{63}\nu_Q = \alpha + \beta \Delta_d , \, \alpha = -73 \, \, \rm{MHz}, \, \beta
= 200 \, \, \rm{MHz}.
\label{eq:nuqdeltad}
\end{equation}

In the same figure we include data obtained from the various materials plotting 
them at the corresponding theoretical values $\Delta_d$. The experimental 
values for $^{63}\nu_Q$ are very accurate but there is of course an uncertainty
in the calculated $\Delta_d$. Furthermore, we expect the theoretical values
for $V_{zz}$ to be less reliable for small $V_{zz}$ due to the cancellations 
of the various contributions shown in~\ref{subsec:contr}. The more reliable  
calculations for the compounds with large $V_{zz}$ (La$_2$CuO$_4$, \ybco{7} 
and \ybc ) systematically predict quadrupole frequencies that are about 
15 \% lower than the experimental ones. The dashed straight line
corresponds to 
\begin{equation}
^{63}\nu_Q = \tilde{\alpha} + \tilde{\beta} \Delta_d = f( \alpha + \beta 
\Delta_d).
\label{eq:exnuqdeltad}
\end{equation}
with $f = 1.2 = 1/0.83 $. This disagreement between calculations and experiments
could be due to a systematic error in the theoretical determination of
$V_{zz}$ or due to a higher value of $^{63}Q$ than assumed or due to both.

It should also be noted that the NQR frequencies $^{63}\nu_Q$ depend on temperature as has been reported in detail by Matsumura {\it et al.} [\onlinecite{matsumura}] for \lacuo. According to the cluster calculations with variable lattice parameter $a$ (see IV F) the general increase of $^{63}V_{zz}$ with temperature in the paramagnetic region is mainly due to the lattice expansion.

Owing to the orthorhombic structure, the EFG at the planar Cu(2) is not 
axially symmetric in the compounds \ybco7 and \ybc\ but shows a slight 
anisotropy since $V_{xx} \ne V_{yy}$. The calculated anisotropy parameters
$\eta = | V_{xx} - V_{yy} | / | V_{zz} |$ are $\eta = 0.047$ and 0.035,
respectively.

The experimentally observed anisotropies have been reviewed by 
Brinkmann\cite{Brinkmann}.
For the planar copper nuclei, they are somewhat smaller than our theoretical 
values. Of more importance, however, are the large $\eta$ values (slightly 
below 1) measured for the chain copper, Cu(1), which are quite unexpected 
since Cu(1) is not at a crystallographic position that would imply $\eta = 1$
by symmetry arguments. An evaluation of the EFG at Cu(1) by cluster methods 
requires that at least the two nearest neighboring planar copper atoms in the
CuO$_2$-planes above and below the Cu(1) site are considered. We have 
previously performed~\cite{bib:huesser1998,bib:renold2001} such calculations 
for a Cu$_3$O$_{12}$ cluster representative of \ybco7. The 
calculated~\cite{bib:renold2001} EFG values for 
Cu(1) are $V_{xx}=0.601$, $V_{yy}=-0.603$ and 
$V_{zz}=-0.002$ a.u. which should be compared with the experimental
ones $V_{xx}=0.767$, $V_{yy}=-0.773$ and                     
$V_{zz}= 0.006$ a.u. as obtained from the measured frequencies with the above
mentioned quadrupole moment $^{63}Q$. The theoretical values produce an asymmetry parameter in complete agreement with the data whereas the absolute value are again about 20 \% smaller than the measured ones.

Recently, Kanigel and Keren\cite{KanigelKeren} have reported NMR measurements
for a series of fully enriched 
(Ca$_{0.1}$La$_{0.9}$)(Ba$_{1.65}$La$_{0.35}$)Cu$_3$O$_y$ powder samples where
doping can vary across the full range from the very underdoped to the extreme
overdoped. They determined the nuclear quadrupole frequency from the four 
peaks of the powder spectra and obtained a convex curve of $\nu_Q$ versus 
doping level which looks very similar to the one (triangles) shown in 
Fig.~\ref{fig:doping}. They used then also a simulation program to account for
asymmetric peaks which provides an increase of $\nu_Q$ with doping in the
underdoped region but a saturation at the overdoped site.

%*****************************************************************************
%*****************************************************************************
\section{Magnetic hyperfine fields}
\label{sec:hyp}
%*****************************************************************************
%*****************************************************************************

\subsection{Theoretical determination}
\label{sec:hypa}
The hyperfine spin Hamiltonians for copper and oxygen nuclei in the CuO$_2$ planes of the cuprates are given by:

\begin{equation}
^{63}H^{\textit{\footnotesize{hf}}}_i = \mathbf I_i \cdot \mathbb A \cdot \mathbf S_i + \sum_{j \in NN} \mathbf I_i \cdot \mathbb B \cdot \mathbf S_j
\end{equation}
and
\begin{equation}
^{17}H^{\textit{\footnotesize{hf}}}_i = \sum_{j \in NN} \mathbf I_i \cdot \mathbb C \cdot \mathbf S_j.
\end{equation}

\noindent
For copper, the hyperfine interaction contains an anisotropic on-site term,
$\mathbb A$, and a transferred term, $\mathbb B$, from the four nearest copper
neighbors, which is usually taken to be isotropic, since it is assumed to
consist only of a contact term. We will, however, retain the tensorial notation of the transferred field, since we will also be able to calculate a dipolar part of the transferred interaction.
For oxygen, the hyperfine interaction $\mathbb C$ is with the two nearest 
copper neighbors.

We have already demonstrated in Refs.~[\onlinecite{bib:huesser2000,bib:renold2001}] that for \lacuo\ and \ybco 7 it is justified to neglect transferred interactions from further distant copper neighbors, both in the case of copper and oxygen.

On a first-principles level, the hyperfine interactions are basically well known, they consist of an isotropic hyperfine density, $D$, a dipolar interaction, $T^{ij}$, and a spin-orbit interaction term. The core polarization is given by the spin density at the nuclear site $R$ and is evaluated as follows:

\begin{equation}
D(\mathbf R) = \frac{8 \pi}3 ( \rho^\uparrow (\mathbf R) - \rho^\downarrow (\mathbf R)).
\end{equation}
If it mainly originates from singly occupied $s$-electrons, it is called Fermi contact. In contrast, if it is due to a doubly occupied $s$-state which is polarized through the spin of other electrons at the same atom or on remote atoms, it is called core polarization. Since in our case, both contributions are present, we prefer to call $D$ just the isotropic hyperfine density.

The dipolar interaction is evaluated as

\begin{equation}
T_{ij}(\mathbf R) = \int d^3r (\rho^\uparrow (\mathbf r)- \rho^\downarrow(\mathbf r))\Delta_{ij}^{\mathbf R}(\mathbf r)
\end{equation}
with

\begin{equation}
\Delta_{ij}^{\mathbf R}(\mathbf r)=\frac{3(r_i-R_i) (r_j-R_j) - \delta_{ij} \abs{\mathbf r -\mathbf R}^2}{\abs{\mathbf r -\mathbf R}^5}.
\end{equation}

\begin{figure}[htb]
\includegraphics[width=0.95\columnwidth]{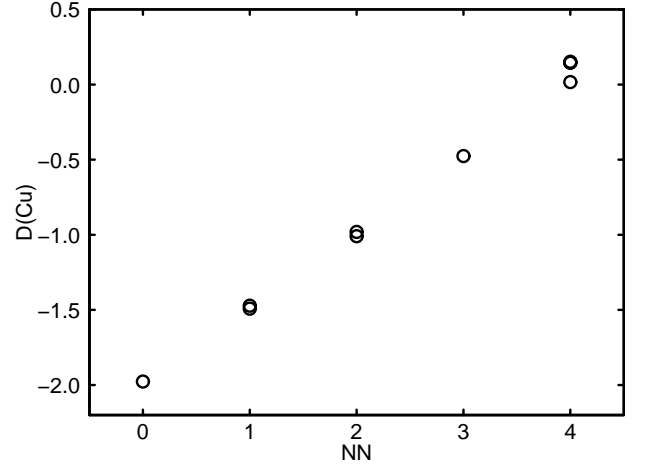}
\caption{Plot of the isotropic hyperfine density recorded at different copper sites in clusters of various sizes for \ybco 6 against the number of nearest copper neighbors at that specific site.}
\label{fig:d_vs_nn}
\end{figure}

The estimation of the spin-orbit interaction will be discussed later.
The above equations yield the total hyperfine fields and still have to be split into on-site and transferred contributions.
To exemplify this splitting we first focus on the clusters with maximal multiplicity. In our clusters for \ybco 6 we find copper atoms with no nearest neighbors (in a small Cu$_1$ cluster), with one (the corner copper in the largest Cu$_{13}$ cluster), with two (e.g. the corner copper in the intermediate Cu$_9$ cluster), with three (the edge copper in the Cu$_{13}$ cluster) and four nearest neighbors. We plot in Fig.~\ref{fig:d_vs_nn} the value of the isotropic hyperfine density against the number $N$ of nearest copper neighbors and find a linear dependence of $D$ on the number of nearest copper neighbors. The value of $D$ when there is no nearest copper neighbor present, is then the on-site contribution, and the slope of the fitted straight line in Fig.~\ref{fig:d_vs_nn} is the transferred contribution per copper neighbor according to the ansatz

\begin{equation}
D = a_{iso} + N b_{iso}.
\label{eq:simpleansatz}
\end{equation}

\noindent
The numerical values for these two contributions are $a_{iso}=-2.00$~a$_B^{-3}$ and $b_{iso}=0.52$~a$_B^{-3}$. In the very same way, one can obtain values for the on-site and transferred part of the dipolar interaction. The $z$-components are $a_{dip}^{\parallel}=-3.38$~a$_B^{-3}$ and $b_{dip}^{\parallel}=0.06$~a$_B^{-3}$. It should be emphasized that in our quantum-chemical calculations the on-site and the transferred terms are highly connected and the linear dependence of $D$ extends from zero NN up to four NN.

By a slight generalization of the ansatz~(\ref{eq:simpleansatz}) it is also possible to include results from clusters with lower multiplicities in the determination of on-site and hyperfine fields. We write

\begin{equation}
D(\textrm{Cu}_i) = \alpha_{iso} \rho_s (\textrm{Cu}_i) + \beta_{iso} \sum_{j \in NN} \rho_s (\textrm{Cu}_j)
\end{equation}
and similarly
\begin{equation}
T^{zz}(\textrm{Cu}_i) = \alpha_{dip}^z \rho_s (\textrm{Cu}_i) + \beta_{dip}^z \sum_{j \in NN} \rho_s (\textrm{Cu}_j).
\end{equation}

For the hyperfine fields at the oxygen site, we make a completely analogous ansatz:

\begin{equation}
D(\textrm{O}_i) = \gamma_{iso} \sum_{j \in NN} \rho_s (\textrm{Cu}_j)
\end{equation}
and
\begin{equation}
T^{zz}(\textrm{O}_i) = \gamma_{dip}^z \sum_{j \in NN} \rho_s (\textrm{Cu}_j).
\end{equation}

The connection between the fitting parameters $\alpha$ through $\gamma$ and
the actual hyperfine parameters is then given by scaling with the expected
spin density in the infinite cluster for which we find a good estimate in the
center of the largest Cu$_{13}$ clusters with ferromagnetic spin multiplicity for each substance, i.e. $a_{iso} = \rho_s(\textrm{Cu}) \alpha_{iso}$ etc..

\begin{figure}[htb]
\includegraphics[width=0.95\columnwidth]{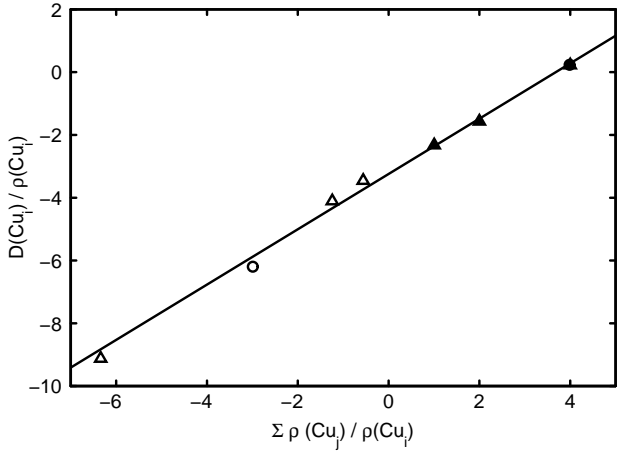}
\caption{Plot of $D(\textrm{Cu}_i)/\rho_s (\textrm{Cu}_i)$ against $\sum_{j \in NN} \rho_s (\textrm{Cu}_j)/\rho_s (\textrm{Cu}_i)$ for \ybco 6. Closed symbols originate from the cluster with ferromagnetic multiplicity $M=14$ and open symbols denote results for antiferromagnetic multiplicity $M=6$. Circles are for values where Cu$_i$ is in the center of the cluster whereas points for off-center coppers are plotted with triangles.}
\label{fig:d_vs_rho}
\end{figure}

The quality of the ansatz can be estimated by suitably chosen plots as e.g. a plot of $D(\textrm{Cu}_i)/\rho_s (\textrm{Cu}_i)$ against $\sum_{j \in NN} \rho_s (\textrm{Cu}_j)/\rho_s (\textrm{Cu}_i)$. (see Fig.~\ref{fig:d_vs_rho}). From the straight line, the fitting parameters $\alpha_{iso}$ and $\beta_{iso}$ are determined and scaled with the expected spin density in the infinite cluster -- as noted above -- to get the hyperfine parameters $a_{iso} = -1.94$~a$_B^{-3}$ and $b_{iso}= 0.77$~a$_B^{-3}$. The corresponding values for the dipolar hyperfine coupling are $a_{dip}^{\parallel}= -3.55$~a$_B^{-3}$ and $b_{dip}^{\parallel}=0.08$~a$_B^{-3}$.
These values are very similar to the ones obtained with the simpler ansatz which shows that the two different kinds of ansatz yield effectively the same results. Because in the second ansatz more clusters are included we quote in the following these results.

\begin{table}[htb]
\begin{tabular}{lrrrr}
                        & \lacuo  & \ybco 6 & \ybco 7 & \ybc\ \\ \hline
$a_{iso}$               & $-$1.94 & $-$2.09    & $-$2.03    & $-$2.08 \\
$b_{iso}$               &    0.77 &    0.57    &    0.50    &    0.55 \\
$a_{dip}^{\parallel}$   & $-$3.55 & $-$3.38    & $-$3.40    & $-$3.38 \\
$b_{dip}^{\parallel}$   &   0.08 &   0.05    &   0.06    &   0.06 \\ \hline
$c_{iso}$               &    0.64 &    0.63    &    0.62    &    0.60 \\
$c_{dip}^{\parallel}$   &    0.40 &    0.40    &    0.39    &    0.42 \\ 
$c_{dip}^{\perp}$   &  $-0.20$&  $-0.21$   &  $-0.21$   &  $-0.22$\\
$c_{dip}^{zz}$   &  $-0.20$&  $-0.19$   &  $-0.18$   &  $-0.20$\\\hline
\end{tabular}
\caption{Theoretical values for the hyperfine parameters at the planar copper and oxygen sites in the four substances determined with the more sophisticated ansatz II using Cu$_{13}$ clusters with ferromagnetic and antiferromagnetic spin arrangement. All values are given in units of a$_B^{-3}$.}
\label{tbl:hyperfinecompilation}
\end{table}

In Table~\ref{tbl:hyperfinecompilation} all hyperfine parameters as determined using the ansatz II can be found. We note that for the oxygen 
$c_{dip}^{\parallel}$ refers to the direction along the bond between the two NN
coppers, $c_{dip}^{\perp}$ is perpendicular to the bond but still in the 
CuO$_2$ plane, while $c_{dip}^{zz}$ denotes the direction perpendicular to the
plane.

The contribution to the hyperfine fields that originate from spin-orbit 
coupling $a^{\alpha}_{so}$ 
are expected to be small in the case of oxygen. For copper, however, they
are of the same order of magnitude as $a_{iso}$ and $a_{dip}$.
We are not  in a position to determine them at the same level of accuracy as $a_{iso}$ 
and $a_{dip}$ and are thus forced to rely on a reasonable estimate. 
In the frame of an atomic picture with a single missing electron in the $3d_{x^2-y^2}$
orbital the dipolar and spin orbit hyperfine interactions are given by
(Ref.~[\onlinecite{bib:bleaney}])
\begin{equation}
a_{dip}^{\parallel} = - \frac{4}7  \expect{r^{-3}} , \, 
a_{so}^{\parallel} = - \frac{62}7 k \expect{r^{-3}}\, \textrm{and} \, a_{so}^{\perp} = -\frac{11}7 k \expect{r^{-3}}.
\label{spinorbit}
\end{equation}
(A value of $k = - 0.044$ was estimated in Ref.~[\onlinecite{bib:monien1990}]).
In a molecule or solid where the missing electron spends some time on the 
oxygen ligands these expressions have to be modified. A simple modification is 
to replace the expression for $a_{dip}^{\parallel}$ by  multiplying it with 
$2 - N_{x^2-y^2}$ $a^{-3}_B$.
Using the values $\expect{r^{-3}} = 8.042$ (Sec. IVD) and $N_{x^2-y^2} = 1.363$
(Table V) we obtain for La$_2$CuO$_4$ $a_{dip}^{\parallel} = -2.93$
which is reasonably close to the directly determined value of $-3.55$ since
actually the spin density should be used instead of the charge density.

For the estimation of the spin-orbit contributions we thus assume that the 
relations (\ref{spinorbit})
still hold in the cluster such that $a_{so}^{\parallel}$ =
$15.5 k a_{dip}^{\parallel}$ and $a_{so}^{\perp} = 2.75 k a_{dip}^{\parallel}$.
The resulting values are given  in Table~\ref{tbl:spinorbit}.

In Table~\ref{tbl:spintotal} we collect the calculated values for the total hyperfine parameters
expressed in terms of densities and also in terms of interaction energies which
we denote by capital letters. The latter are defined by $^{63}A_{tot}^{\alpha} =
\hbar \gamma_e \hbar ^{63}\gamma a_{tot}^{\alpha}$ and similarly for the $^{63}B$ and $^{17}C$ and (as the notation) indicates depend on the particular isotope.  

\begin{table}[htb]
\begin{tabular}{lrrrr} \hline
                      & \lacuo    & \ybco 6 & \ybco 7 & \ybc\ \\ \hline
$a_{so}^{\parallel}$  &    2.43 &    2.32 &    2.27 &    2.28 \\
$a_{so}^{\perp}$      &    0.43 &    0.41 &    0.40 &    0.40 \\ \hline
\end{tabular}
\caption{Estimation for spin-orbit contributions to the hyperfine fields at 
the copper site in the various substances. All values are given in atomic 
units.}
\label{tbl:spinorbit}
\end{table}

\begin{table}[htb]
%\begin{tabular}{lrrrrc|lrrrr} \hline
% & LACUO & YBCO6 & YBCO7 & YBCO8&\,\,&& LACUO &YBCO6 & YBCO7 & YBCO8  \\ \hline
%$a_{tot}^{\parallel}$  & $-3.06$ & $-3.15$ & $-3.16$ & $-3.18$ &&
%\,$^{63}A_{tot}^{\parallel}$  & $-1.79$ & $-1.84$ & $-1.85$ & $-1.86$ \\
%$a_{tot}^{\perp}$      &    0.26 & 0.01    & 0.07    & 0.01&&
%\,$^{63}A_{tot}^{\perp}$      &    0.15 & 0.01    & 0.04    & 0.01     \\
%$b_{tot}^{\parallel}$  & 0.85 & 0.62 & 0.56 & 0.61 &&
%\,$^{63}B_{tot}^{\parallel}$  & 0.50 & 0.36 & 0.33 & 0.36 \\
%$b_{tot}^{\perp}$      & 0.73 & 0.52 & 0.47 & 0.52 &&
%\,$^{63}B_{tot}^{\perp}$      & 0.43 & 0.30 & 0.27 & 0.30 \\  
%$c_{tot}^{\parallel}$  &    1.04 & 1.03    & 1.01    & 1.02 && 
%\,$^{17}C_{tot}^{\parallel}$  & 0.31 & 0.31 & 0.30 & 0.30  \\  
%$c_{tot}^{\perp}$      &    0.44 & 0.42    & 0.41    & 0.38 &&
%\,$^{17}C_{tot}^{\perp}$      & 0.13 & 0.13 & 0.12 & 0.11   \\
%$c_{tot}^c$         &    0.44 & 0.44    & 0.44    & 0.40 &&   
%\,$^{17}C_{tot}^c$         & 0.13 & 0.13 & 0.13 & 0.12 \\\hline
\begin{tabular}{lrrrr} \hline
 & \lacuo\ &\ybco6 & \ybco7 & \ybc\ \\ \hline
$a_{tot}^{\parallel}$  & $-3.06$ & $-3.15$ & $-3.16$ & $-3.18$ \\
$a_{tot}^{\perp}$      &    0.26 & 0.01    & 0.07    & 0.01 \\
$b_{tot}^{\parallel}$  & 0.85 & 0.62 & 0.56 & 0.61 \\
$b_{tot}^{\perp}$      & 0.73 & 0.52 & 0.47 & 0.52 \\
$c_{tot}^{\parallel}$  &    1.04 & 1.03    & 1.01    & 1.02 \\
$c_{tot}^{\perp}$      &    0.44 & 0.42    & 0.41    & 0.38 \\
$c_{tot}^c$         &    0.44 & 0.44    & 0.44    & 0.40 \\\hline
% & \lacuo\ &\ybco6 & \ybco7 & \ybc\ \\ \hline
\,$^{63}A_{tot}^{\parallel}$  & $-1.79$ & $-1.84$ & $-1.85$ & $-1.86$ \\
\,$^{63}A_{tot}^{\perp}$      &    0.15 & 0.01    & 0.04    & 0.01     \\
\,$^{63}B_{tot}^{\parallel}$  & 0.50 & 0.36 & 0.33 & 0.36 \\
\,$^{63}B_{tot}^{\perp}$      & 0.43 & 0.30 & 0.27 & 0.30 \\
\,$^{17}C_{tot}^{\parallel}$  & 0.31 & 0.31 & 0.30 & 0.30  \\
\,$^{17}C_{tot}^{\perp}$      & 0.13 & 0.13 & 0.12 & 0.11   \\
\,$^{17}C_{tot}^c$         & 0.13 & 0.13 & 0.13 & 0.12 \\\hline
\end{tabular}
\caption{Contributions to the hyperfine fields at
the copper site ($a$ and $b$) and at the oxygen site ($c$) in the 
various compounds in a$_B^{-3}$ ($a,b$ and $c$) and in $\mu$eV ($^{63}A$, 
$^{63}B$ and $^{17}C$).
$^{63}A/a$ and $^{63}B/b$ = 0.5844 $\mu$eVa$_B^3$ and $^{17}C/c$ =
0.2988 $\mu$eVa$_B^3$.}
\label{tbl:spintotal}
\end{table}
\subsection{Comparison with experiments}

Although experiments cannot determine on-site and transferred hyperfine fields separately it is possible to extract various combinations of on-site and transferred fields using different experimental set-ups.

One constraint is the relation $a_{tot}^{\parallel} + 4b_{tot}^{\parallel}=0$ 
which explains that the NMR spin shift measured with the field in $c$-direction
does not change below the superconduction transition temperature $T_c$ in 
contrast to the spin shift measured with the field perpendicular to the 
$c$-axis. We postpone a comparison with our theoretical values to 
Sec.~\ref{sec:theodet}.

A second experimental determination of a combination of on-site and transferred hyperfine fields is made possible through measurements of the NMR resonance frequency, $^{63}\nu_L$, of the copper nuclei in the pure parent compounds La$_2$CuO$_4$ and YBa$_2$Cu$_4$O$_6$ which are in the antiferromagnetic state. This determines the local magnetic field $^{63}\gamma H_{loc} = 2\pi ^{63}\nu_L$ which is commonly expressed as the corresponding hyperfine field in units of an effective electronic magnetic moment $\mu_B^{\textit{\footnotesize{eff}}} \simeq 0.66 \mu_B$ as $H_{loc} = \abs{a_{tot}^{\perp} - 4b_{tot}^{\perp}}\mu_B^{\textit{\footnotesize{eff}}}$.

For \lacuo\ the theoretical value for $\Delta \equiv \abs{a_{tot}^{\perp} - 4b_{tot}^{\perp}}$ is 2.66 whereas the measured\cite{bib:tsuda1988} frequency of 93.85 MHz corresponds to $\Delta = 2.01$. For \ybco 6 we get $\Delta = 1.90$ in good agreement with the experiment\cite{bib:yasuoka1988} (89.89 MHz) which leads to $\Delta = 1.92$.

Sometimes the measured anisotropy between the copper spin-lattice relaxation 
times $^{63}T_{1,\parallel} / ^{63}T_{1,\perp}$ has also been used to extract 
information about the hyperfine coupling constants. This, however, is only
possible for the two extreme conditions of totally antiferromagnetic 
correlations (which is never achieved for the doped samples) or of no
correlations, which would require measurements at very high temperatures. 

\begin{table}[htb]
\begin{tabular}{lrrrrr} \hline
 & $a)$ & $b)$ & $c)$ & $d)$ & $e)$  \\ \hline
$^{63}A_{tot}^{\parallel}$  & $-1.85$ & $-1.86$ & $-1.76$ & $-1.61$ &$-0.94$ \\
$^{63}A_{tot}^{\perp}$      & 0.04    & 0.23 & $-0.10$& 0.29 & 0.17  \\
$^{63}B_{iso}$  & 0.29 & 0.34 & 0.41 & 0.40 & 0.23 \\ \hline
\end{tabular}
\caption{Contributions to the hyperfine fields at the copper site in \ybco7. 
$a)$ this work, $b)$ Ref.~[\protect \onlinecite{bib:monien1990}],
$c)$ Ref.~[\protect \onlinecite{bib:walstedt1990a}], 
$d)$ Ref.~[\protect \onlinecite{bib:zha}], and
$e)$ Ref.~[\protect \onlinecite{bib:nandor}]. } 
\label{tbl:hyperexp}
\end{table}

In Table~\ref{tbl:hyperexp} we compare our values for the hyperfine interaction
energies with those published by several authors. An inspection shows that the 
deviations are not large but sufficiently strong to render further 
interpretations questionable. In particular,
until more reliable values for the influence of the spin-orbit coupling on the hyperfine fields are known, it is prohibitive to make more precise statements.

%*****************************************************************
%*****************************************************************
\section{Chemical shieldings and Paramagnetic Field Modifications}
\label{sec:chem}
%*****************************************************************
%*****************************************************************

%***************************
\subsection{General remarks}
%***************************

Very early after the discovery of the high temperature superconductors, numerous measurements of Knight shifts at various nuclei have been performed. For optimally doped \ybco7, the Knight shift $\Kk{63}{\perp}$ of the planar copper for the field applied perpendicular to the $c$-axis is temperature independent above $T_c$ and drops below $T_c$ with decreasing temperature to $\Kk{63}{\perp}(T=0)$. The behavior above $T_c$ is to be expected for a temperature independent Pauli spin susceptibility. The reduction of the Knight shift below $T_c$ was explained by the formation of Cooper pairs which are -- due to their vanishing total spin -- not available for polarization by a magnetic field and therefore do not contribute to the Knight shift. At zero temperatures, all charge carriers were assumed to be bound in Cooper pairs and the remaining Knight shift $\Kk{63}{\perp}(T=0)$ was attributed to the temperature and doping independent chemical shift. The temperature dependence of $\Kk{63}{\perp}(T)$ in the superconducting state depends on the symmetry of the pairing state and most NMR experiments were better explained by $d$-wave pairing.
(For underdoped materials, the decline of the Knight shift $\Kk{63}{\perp}(T)$ with lowering temperatures sets in already at temperatures above $T_c$ which has attributed to the opening of a spin pseudogap.) 

The copper Knight shift $\Kk{63}{\parallel}(T)$ with the applied field along the crystallographic $c$ axis, however, is constant over the whole temperature range, i.e. it is completely unaffected by the superconducting transition. Therefore, it was concluded that all of the measured Knight shifts are of chemical origin. The vanishing spin part of the Knight shifts was then explained by an accidental cancellation of the on-site and transferred hyperfine fields. As already mentioned in Sec.~\ref{sec:hyp} our calculations of on-site and transferred hyperfine fields in substances of the La and Y families do not support this cancellation.
In this section we will, in addition, give further evidence that the above sketched explanation of the Knight shifts in cuprates needs careful revision based on first-principles calculations of chemical shifts at the planar copper nuclei in La$_2$CuO$_4$, YBa$_2$Cu$_3$O$_6$, and YBa$_2$Cu$_3$O$_7$. Before we report on the results of these calculations, we point out that copper Knight shift measurements have long been misinterpreted due to a wrong assumption on the magnetic properties of the reference substance. As a remedy we introduce a new term, the paramagnetic field modification.

%%%%%%%%%%%%%%%%%%%%%%%%%%%%%%%%%%%%%%%%%%%%%%%%%%%%%%%%%%%%%%%%%%%%%%%%
\subsection{The role of the reference substance}
%%%%%%%%%%%%%%%%%%%%%%%%%%%%%%%%%%%%%%%%%%%%%%%%%%%%%%%%%%%%%%%%%%%%%%%%

Knight shift measurements are in fact measurements of differences in resonance frequencies of a nuclear species $k$, in a target substance ($t$) and in a reference substance ($r$).
\begin{equation}
\K{k}{}{}(t-r) = \frac{^k\nu (t) - \mbox{$^k\nu$} (r)}{\mbox{$^k\nu$} (r)}.
\end{equation}
For copper Knight shift measurements the most often used reference substance is the monovalent CuCl. The Knight shifts are made up of two parts, a temperature independent chemical shift, $\K{k}{L}{}$, and a spin shift, $\K{k}{s}{}$. In this section we consider the contribution of the chemical shift.

For a theoretical determination of chemical shifts we need to know the chemical shieldings, $\s{}{}{}$, in the target and the reference substance. The connection to the measured chemical shift, $\K{k}{L}{ii}(t-r)$, is given by
\begin{equation}
\K{k}{L}{ii}(t-r) = \frac{{\s{k}{}{ii}(r)} - {\s{k}{}{ii}(t)}}{1+{\s{k}{}{ii}(r)}} \simeq {\s{k}{}{ii}(r)} - {\s{k}{}{ii}(t)},
\end{equation}
or, with the separation of $\s{k}{}{ii}$ into diamagnetic $(d)$ and paramagnetic $(p)$ parts of the shieldings, by
\begin{equation}
\K{k}{L}{ii}(t-r) = {\s{k}{d}{ii}(r)} + {\s{k}{p}{ii}(r)} - {\s{k}{d}{ii}(t)} - {\s{k}{p}{ii}(t)}.
\label{eq:related}
\end{equation}
As we have shown in Ref.~[\onlinecite{bib:renold2003}] the differences between $\s{63}{d}{ii}(r)$ and $\s{63}{d}{ii}(t)$ are negligible and we can therefore write
\begin{equation}
\K{63}{L}{ii}(t-r) = {\s{63}{p}{ii}(r)}  - {\s{63}{p}{ii}(t)}.
\label{eq:para}
\end{equation}
It has long been assumed, that the paramagnetic contribution of the shielding in the target substance, CuCl, is small. We have shown, however, ($i$) by a direct quantum chemical calculation and ($ii$) by referring to measurements of the chemical shieldings using atomic beam techniques, that $\s{63}{p}{ii}($CuCl$) = 1500$~ppm. In view of typical chemical shieldings at copper nuclei in cuprates (see Sec.~\ref{sec:theodet}) this contribution is sizeable and cannot be neglected.

The quantities of interest in Eq.~(\ref{eq:para}) are, of course, not the chemical shifts, $\K{63}{L}{ii}(t-r)$, but rather the contributions of the target, i.e. $-\s{63}{p}{ii}(t)$. To avoid misinterpretations we find it most convenient to introduce here a new quantity, the paramagnetic field modification:
\begin{equation}
\Kbar{63}{L}{ii}(t) \equiv -\s{63}{p}{ii}(t) = {\K{63}{L}{ii}(t-r)} - {\s{63}{p}{ii}(r)}. 
\label{kbar}
\end{equation}
It is important to note that this paramagnetic field modification $\Kbar{63}{L}{ii}(t)$ is independent of the reference substance.

%%%%%%%%%%%%%%%%%%%%%%%%%%%%%%%%%%%%%%%%%%%%%%%%%%%%%%%%%%%%%%%%%%%%%%%%%%%%%%%%%%%%%%%%%%%
\subsection{Theoretical determination of chemical shifts and comparison to experiment}
\label{sec:theodet}
%%%%%%%%%%%%%%%%%%%%%%%%%%%%%%%%%%%%%%%%%%%%%%%%%%%%%%%%%%%%%%%%%%%%%%%%%%%%%%%%%%%%%%%%%%%

For a detailed description about the determination of chemical shifts in the framework of the cluster technique, we refer the reader to Refs.~[\onlinecite{bib:renold2003,bib:renolddiss}]. Here we just mention the important steps.

Calculations have been performed for La$_2$CuO$_4$, YBa$_2$Cu$_4$O$_6$, and YBa$_2$Cu$_3$O$_7$ using clusters with five copper atoms in the plane. Larger clusters have not been employed systematically since the determination of the shielding constants with the large basis sets that are required for accurate calculations are extremely time consuming. Test calculations in Cu$_9$ clusters for La$_2$CuO$_4$ have shown, however, that the results do not change upon enlarging the clusters. 

\begin{figure}
\begin{center}
\includegraphics[width=1.0\columnwidth]{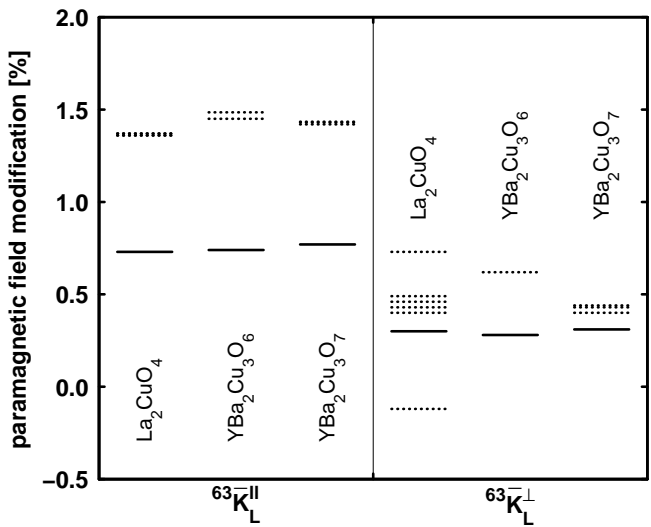}
\caption{Paramagnetic field modifications at the planar copper site in the considered substances both for fields along the $c$ axis, $\Kbar{63}{L}{\parallel}$, (left panel) and for fields in the plane, $\Kbar{63}{L}{\perp}$ (right panel). Results obtained theoretically (experimentally) are displayed with solid (dashed) bars. The experimental data are taken from Ref.~[\protect \onlinecite{bib:ohsugi}] (La$_{2-x}$Sr$_x$CuO$_4$), from Refs.~[\protect \onlinecite{bib:mali1991,bib:pozzi1999}] (YBa$_2$CuO$_6$) and from Refs.~[\protect \onlinecite{bib:barrett1991,bib:walstedt1990,bib:takigawa1991}] (YBa$_2$CuO$_{7-\delta}$).}
\label{fig:comp_th_exp_cu_planar}
\end{center}
\end{figure}

In Fig.~\ref{fig:comp_th_exp_cu_planar} theoretical results for $\Kbar{63}{L}{\parallel}$ ($\Kbar{63}{L}{\perp}$) are displayed with solid bars in the left (right) panel. The dotted bars denote results reported from various experiments.
It is observed that for the applied field in the plane (right panel of Fig.~\ref{fig:comp_th_exp_cu_planar}) the theoretical values are in general slightly lower than the values obtained from experiments. For fields parallel to the $c$ axis, theory predicts values for $\Kbar{63}{L}{\parallel}$ (left panel) that are only about half as large as the experimental results. We find, however, that the paramagnetic field modifications at the planar copper sites hardly depend on the specific cuprate. 

We have, at present, no explanation for the discrepancy between theory and experiment. We would like to point out, however, that it was shown in Refs.~[\onlinecite{bib:zheng99_1248,bib:zheng00_tlperp}] that the measured Knight shifts below $T_c$ are field dependent and drop when reducing the applied magnetic field $H$. The theoretical calculations, of course, are in the limit of $H \rightarrow 0$. Furthermore, it is also possible that impurities induce a finite density of states at $T=0$, as was proposed in Ref.~[\onlinecite{bib:ohsugi}]. Both of the above two ideas imply that the measured Knight shifts at $T=0$ are not entirely of chemical origin but also have contributions from spin degrees of freedom.

The temperature independence of the copper Knight shift when measured with the field in c-direction is most easily explained by an incidentend cancellation of the on-site and transferred hyperfine fields, i.e. $a_{tot}^{\parallel} +
4 b_{tot}^{\parallel} \approx 0$. In Table X we present the values calculated for
$ 1 + a_{tot}^{\parallel} / 4 b_{tot}^{\parallel}$ for the four substances under consideration. The values are close to zero but differ among the various compounds. For \lacuo, a small positive value of 0.1 is obtained. for \ybco7, however, we obtain $-0.4$. This difference is from the theoretical point of view easily explained by the fact that the transferred hyperfine field $b_{tot}^{\parallel}$ in \ybco7 is smaller than in \lacuo\ due to the buckling of the planar oxygen atoms. It is evident that the calculations for one compound may fail to give the cancellations necessary for the easy explanation of the temperature independence of $^{63}K^{\parallel}$. To explain the behavior of $^{63}K^{\parallel}$ both in \lacuo\ and in the Y-compounds already requires a double coincidence. In addition, measurements on the electron doped material PrLaCeCuO$_4$ by Zheng {\it et al.}~\cite{bib:Zheng} also exhibit a temperature independent $^{63}K^{\parallel}$. In view of the differences in the lattice parameters in all three substances it is extremely intriguing that these cancellations of on-site and transferred hyperfine fields which are basically determined by chemistry, occur.

\begin{table}
\begin{tabular}{lc}
\hline
& 1 + $a_{tot}^{\parallel} / 4 b_{tot}^{\parallel}$ \\
\hline
La$_2$CuO$_4$      &$\, \,\,\,0.105$ \\
YBa$_2$Cu$_3$O$_6$ & $-$0.278 \\
YBa$_2$Cu$_3$O$_7$ & $-$0.402 \\
YBa$_2$Cu$_4$O$_8$ & $-$0.292 \\
\hline
\end{tabular}
\label{tbl:totalhyp}
\caption{Theoretical relations for the total hyperfine fields at the
central copper atom.}
\end{table}

%********************************
%********************************
\section{Summary and Conclusions}
\label{sec:summary}
%********************************
%********************************

We have performed large-scale ab-initio cluster computations of cuprates in order to determine the local electronic structure. The convergence of these local properties with respect to the cluster size is very good. An analysis of the charge and spin distribution in terms of contributions from the various MO and AO reveals distinguished features in all compounds under consideration. First, the copper 3$d_{x^2-y^2}$ AO is occupied by about 1.4 and the 3$d_{3z^2-r^2}$ AO by about 1.9 electrons. The oxygen 2$p_{\sigma}$ AO contains roughly 1.65 electrons. This implies a total of 1.4 intrinsic holes per unit which is compensated by $0.4-0.5$ electrons in the copper 4$s$ AO. These partial occupancies of the non-spherical AO mainly determine the EFG values. The 4$s$ AO is involved in the transferred hyperfine field. Good agreement between the calculated and measured copper EFG is found. The EFG values essentially depend on the differences between the Mulliken populations of the 3$d_{3z^2-r^2}$ and the 3$d_{x^2-y^2}$ AO.

Simulating doping by two different methods shows that all these occupancies smoothly change and the general trends of changing copper EFG with doping level are reproduced. The removal of one electron by a dopant atom in the intra-layer 
induces only half a hole in the CuO$_2$ plane. The other half is in 
out-of-plane orbitals.

Spin-polarized calculations with various spin multiplicities enabled the determination of the antiferromagnetic exchange coupling and the various hyperfine fields. The contribution of the spin-orbit coupling, however, has only been approximately determined. The calculated total hyperfine fields are in rough agreement with those deduced from experiments. The values for the sum $a_{tot}^{\parallel} + 4 b_{tot}^{\parallel}$ are small but differ considerably among the substances under consideration. This in sharp contrast to the requirements set by the temperature independence of the copper Knight-shift $^{63}K^{\parallel}$ observed in very different cuprates. 

We conclude that the out-of-plane orbital 3$d_{3z^2-r^2}$ and the 4$s$ orbital play a more important role than commonly assumed.

\acknowledgments
We express our gratitude to M. Mali, J. Roos and C. P. Slichter for numerous stimulating discussions.
This work was partially supported by the Swiss National Science Foundation.

%%%%%%%%%%%%%%%%%%%%%%%%%%%%%%%%%%%%%%%%%%%%%%%%%%%%%%%%%%%%%%%%%%%%%%%%%%%


\begin{thebibliography}{99}
\bibitem{footnote}{in the present context it is not necessary to distinguish
between the different spin projections.}
\end{thebibliography}
\end{document}